\documentclass[prb,superscriptaddress,floatfix,showpacs,amsmath,amssymb]{revtex4}

\usepackage{graphicx}
\usepackage{dcolumn}
\usepackage{bm}
\usepackage{color}
\usepackage{tabularx}

\newcommand{\lfp}{LiFePO$_4$}
\newcommand{\lmp}{LiMnPO$_4$}
\newcommand{\lcp}{LiCoPO$_4$}
\newcommand{\lnp}{LiNiPO$_4$}

\newcommand{\tn}{$T_{\rm N}$}
\newcommand{\bsf}{$B_{\rm SF}$}

\newcommand{\rk}{\color{black}}

\begin{document}


\title{Exceptional field dependence of antiferromagnetic magnons in LiFePO$_4$}

\author{J.~Werner}
\affiliation{Kirchhoff Institute of Physics, Heidelberg University, INF 227, D-69120 Heidelberg, Germany}
\author{C.~Neef}
\affiliation{Kirchhoff Institute of Physics, Heidelberg University, INF 227, D-69120 Heidelberg, Germany}
\affiliation{Fraunhofer Institute for Systems and Innovation Research ISI, Karlsruhe D-76139, Germany}
\author{C.~Koo}
\affiliation{Kirchhoff Institute of Physics, Heidelberg University, INF 227, D-69120 Heidelberg, Germany}
\author{A.~Ponomaryov}
\affiliation{Dresden High Magnetic Field Laboratory (HLD-EMFL), Helmholtz-Zentrum Dresden Rossendorf, D-01328 Dresden, Germany}
\author{S.~Zvyagin}
\affiliation{Dresden High Magnetic Field Laboratory (HLD-EMFL), Helmholtz-Zentrum Dresden Rossendorf, D-01328 Dresden, Germany}
\author{R.~Klingeler}
\email[Email: ]{klingeler@kip.uni-heidelberg.de}
\affiliation{Kirchhoff Institute of Physics, Heidelberg University, INF 227, D-69120 Heidelberg, Germany}
\affiliation{Centre for Advanced Materials (CAM), Heidelberg University, INF 225, D-69120 Heidelberg, Germany}

\date{\today}
\begin{abstract}
Low-energy magnon excitations in magnetoelectric \lfp\ have been investigated by high-frequency high-field electron spin resonance spectroscopy in magnetic fields up to $B = 58$~T and frequencies up to $f = 745$~GHz. For magnetic fields applied along the easy magnetic axis, the excitation gap softens and vanishes at the spin-flop field of $B_\mathrm{SF} = 32$~T before hardening again at higher fields. In addition, for $B\lesssim B_\mathrm{SF}$ we observe a resonance mode assigned to excitations due to Dzyaloshinskii-Moriya (DM)-interactions, thereby evidencing sizable DM interaction of $\approx 150~\mu$eV in \lfp . Both the magnetisation and the excitations up to high magnetic fields are described in terms of a mean-field theory model which extends recent zero field inelastic neutron scattering results. 
Our results imply that magnetic interactions as well as magnetic anisotropy have a sizeable quadratic ﬁeld dependence which we attribute to signiﬁcant magnetostriction. 
\end{abstract}
\maketitle

\section{Introduction}

There is a variety of experimental techniques to probe and elucidate spin dynamics in magnetically ordered materials, among them inelastic neutron scattering (INS), resonant inelastic X-ray scattering (RIXS), Raman- and terahertz spectroscopy, or high-frequency electron spin resonance spectroscopy (HF-ESR). While probing the same low-temperature magnetic excitations aiming at establishing the magnetic Hamiltonian for a material, the experiments are complementing each other as, e.g., different sample size, resolution, energy and reciprocal space regimes, or magnetic fields are accessible. One illustrative example is BiFeO$_3$ where INS~\cite{jeong2012spin,matsuda2012magnetic,xu2012thermal}, terahertz-spectroscopy~\cite{talbayev2011long,nagel2013terahertz,fishman2015spin}, HF-ESR~\cite{ruette2004magnetic} and Raman~\cite{cazayous2008possible,rovillain2010electric} data have lead to a conclusive view and established the microscopic spin Hamiltonian. Our current work on \lfp\ highlights not only the strength of  antiferromagnetic resonance (AFMR) studies to probe low-energy magnetic excitations but also illustrates how microscopic parameters can change upon application of high magnetic fields, e.g., by magnetostrictive effects.

\begin{figure}	
\includegraphics[width=0.5\columnwidth,clip] {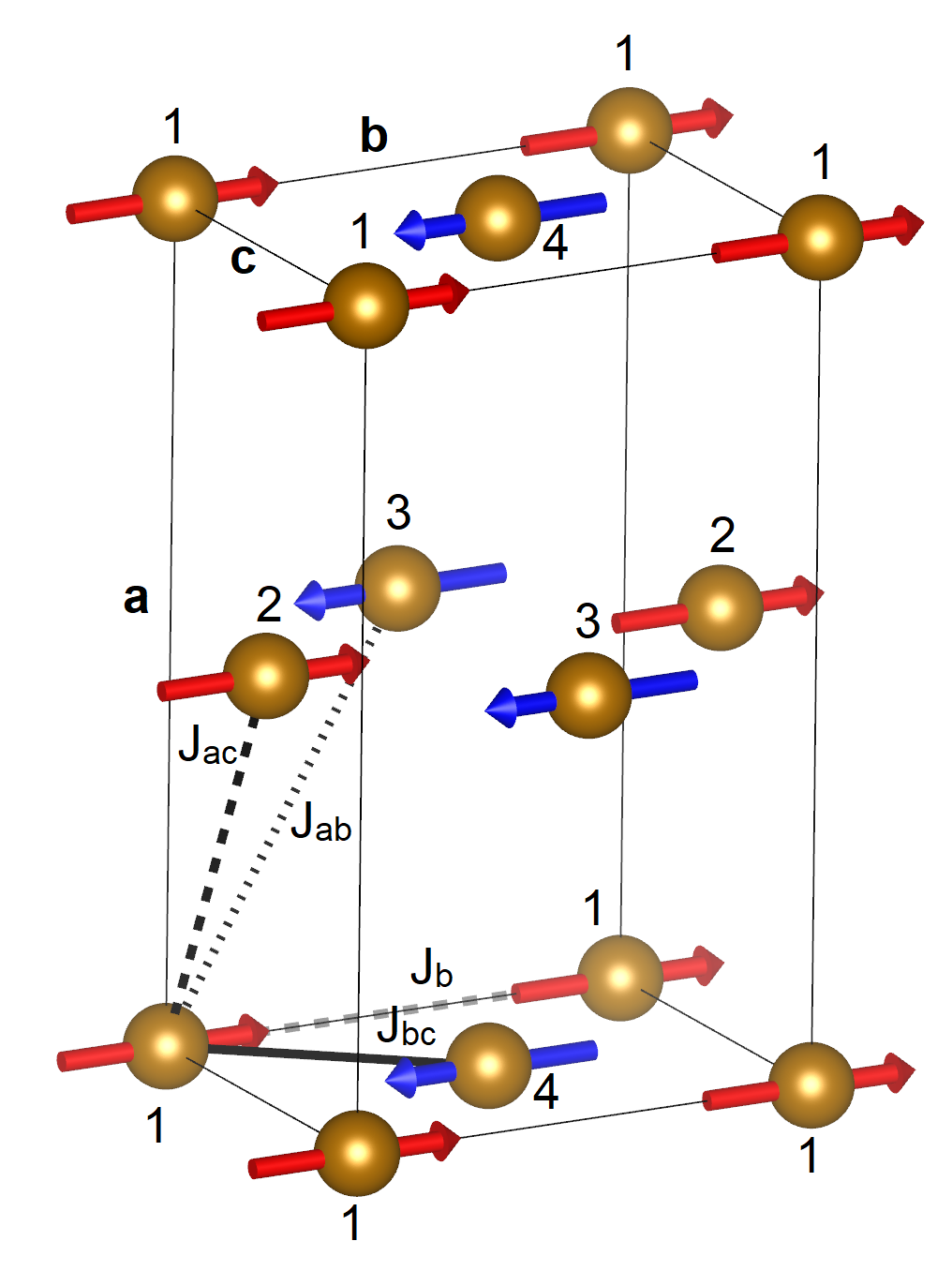}
\caption{Schematic view of magnetic structure of \lfp~at zero field, based on Refs.~\onlinecite{toft2015anomalous} and \onlinecite{yiu2017hybrid}. Numbers illustrate how the Fe-moments are associated with four antiferromagnetic sublattices. $J_{ab}$, $J_{ac}$, $J_{bc}$, and $J_b$ are the leading magnetic interactions. The figure has been generated with VESTA~\cite{momma2011vesta}.}
\label{magStru}
\end{figure}

Due to intriguing magnetoelectric properties~\cite{zimmermann2009anisotropy,rivera1994linear,van2008anisotropy,van2007observation,vaknin2004commensurate,toft2015anomalous}, spin excitations in olivine-type lithium orthophosphates Li$M$PO$_4$ ($M$ = Mn, Fe, Co, Ni) have been measured by various experiments. While for \lmp\ and \lnp , complementing investigations by INS, THz- and Raman-spectroscopy enabled to establishing microscopic spin models, for \lcp\ and \lfp\ comprehensive models are, however, still missing. Among the above mentioned olivine-type phosphates, \lmp\ shows the smallest magnetocrystalline anisotropy as probed by ESR in the paramagnetic phase and around the antiferromagnetic (AFM) ordering temperature~\cite{arvcon2004weak}. Inelastic neutron scattering revealed accordingly small magnon excitation gaps of $\Delta_1 \approx 120$~GHz and $\Delta_2 \approx 160$~GHz in the ordered phase~\cite{li2009antiferromagnetism}. Raman data corroborated a microscopic spin model which together with INS results led to a robust set of magnetic parameters~\cite{calderon2015two}. A similar picture evolves for \lnp\ from Raman and INS data where low energy spin excitation gaps of $\Delta_1 \approx 450$~GHz and $\Delta_2 \approx 1.1$~THz were found~\cite{fomin2002raman,jensen2009anomalous,toft2011high,li2009tweaking}. Infrared absorption spectroscopy confirm the INS results \cite{peedu2019spin} and recent Raman scattering results provide an independent experimental confirmation of previously obtained exchange parameters\cite{rigitano2020raman}. In contrast, a conclusive model for \lcp\ is still under debate which might be due to the strong magnetoelectric effect in this material~\cite{rivera1994linear}. Spin excitations in \lcp\ have been studied extensively both by INS~\cite{vaknin2002weakly,tian2008spin,tian2010neutron} and infrared-absorption spectroscopy~\cite{kocsis2019magnetoelectric,kocsis2018identification}. It was revealed that the excitation gaps in the center of the Brillouin zone are $\Delta_1 \approx 1.1$~THz and $\Delta_2 \approx 1.3$~THz. 
Here, we report on \lfp\ where previous investigations of low energy spin excitations have been performed by Raman spectroscopy~\cite{paraguassu2005phonon} and INS~\cite{li2006spin,yiu2017hybrid,toft2015anomalous}. {\rk The zero-field spin configuration based on Refs.~\onlinecite{toft2015anomalous} and \onlinecite{yiu2017hybrid} is shown in Fig.~\ref{magStru}. While the magnetic moments are pointing mainly along the crystallographic $b$-axis, there is also a small canting which indicates the presence of Dzyaloshinsky-Moriya interaction~\cite{toft2015anomalous}.} The INS reports agree that the relevant exchange interactions are all antiferromagnetic and that the dominant interaction is $J_{bc}$ (see Fig.~\ref{magStru}) but the exact values differ significantly.~\footnote{Note, that $J_b$ does not significantly affect the AFMR model at $B<B_{\rm SF}$ and hence is omitted in the Hamiltonian Eq.~\ref{spinH}.}
Furthermore, the INS results can not explain the static magnetic properties in magnetic fields at low-temperature, as a spin-flop-like transition at $B_\mathrm{SF} = 32$~T is not reproduced by the INS model.

To clarify these issues and to investigate the spin-flop transition in more detail we performed high-frequency/high-field electron spin resonance spectroscopy (HF-ESR) studies on \lfp . Our study reveals the presence of low-energy magnon modes around the spin-ﬂop ﬁeld. Both the softening of these magnon modes towards \bsf\ and the magnetisation are described by means of a mean-field theory model which, while based on the INS model, employs field dependent values of magnetic interaction and magnetic anisotropy. In addition, we find direct evidence for finite DM interactions. The estimated size of the DM interaction is $150 \mu$eV. Measurements at higher temperatures reveal the disappearance of the resonances already well below the antiferromagnetic ordering temperature.

\section{Experimental}

Single crystals of \lfp\ were grown by the high-pressure optical floating-zone method as reported in detail in Ref.~[\onlinecite{neef2017high}]. High-frequency electron spin resonance spectroscopy (HF-ESR) experiments in pulsed fields up to 50 T and magnetization studies up to 58~T were performed at the Dresden High Magnetic Field Laboratory (HLD). The HF-ESR data were obtained using VDI modular transmitters (product of Virginia Diodes Inc., USA) as sub-mm radiation sources and InSb hot-electron bolometer as a radiation detector. {\rk The experiments have been performed at frequencies $f$ between 65 and 1100~GHz. However, at 750~GHz $\leq f \leq$ 1100 GHz, there is strong magnetic field-independent microwave absorption, possibly due to non-magnetic excitation, which does not allow to obtain reliable data in this frequency regime.} Magnetization studies employed a coaxial pick-up coil system; the magnetization data were calibrated using data, obtained in static fields at 5~T (see Ref.~[\onlinecite{werner2019high}]).

\section{Results}

\begin{figure}	
\includegraphics[width=1\columnwidth,clip] {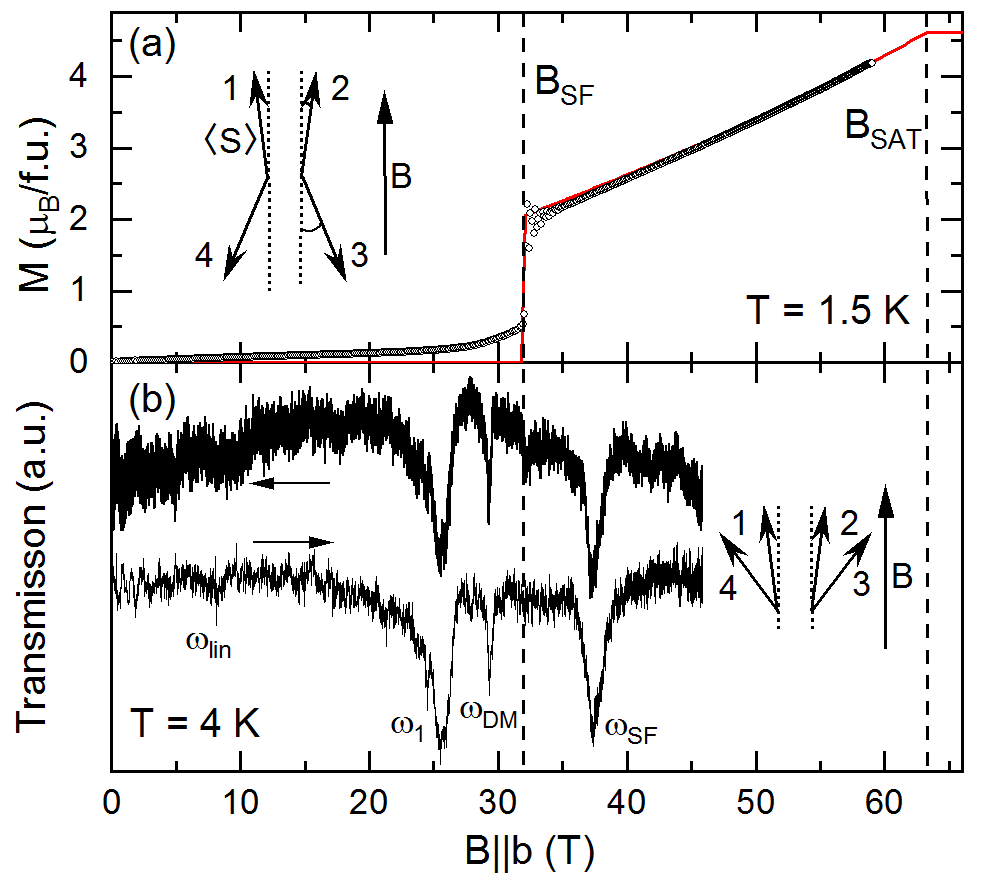}
\caption{(a) Pulsed-field magnetization of \lfp\ with $B||b$ at $T = 1.5$~K~\cite{werner2019high}. Red line is a simulation of the data, see the text. Vertical dashed lines indicate the spin-flop field $B_\mathrm{SF}$ and the saturation field $B_\mathrm{sat}$. (b) Pulsed-field electron spin resonance up-sweep ($\rightarrow$) and down-sweep ($\leftarrow$)  spectra at $T = 4$~K and $f = 708$~GHz. The resonances are labeled with $\omega_\mathrm{lin}$ for the linear resonance branch and $\omega_\mathrm{1,DM,SF}$ (see Fig.~\ref{FH}). Pictograms indicate the magnetic structure respectively the model for $B||b$ (a) at $B<B_{\rm SF}$ (adapted from Ref.~\onlinecite{toft2015anomalous}), and (b) in the spin-flop phase.}
\label{mag}
\end{figure}

{\rk Long-range  antiferromagnetic  order in LiFePO$_4$ single crystals studied at hand~\cite{neef2017high} evolves at \tn\ = 50.0(5)~K with the crystallographic $b$-axis being the magnetic easy axis.} Magnetic fields applied along the magnetic easy axis of \lfp\ leads to a jump in magnetization at around $B_\mathrm{SF} = 32$~T {\rk and a rather linear behavior above the metamagnetic transition, see Fig.~\ref{mag}a. While this behavior strongly reminds on a typical spin-reorientation transition, extrapolating of the high-field} magnetization to zero field does not yield the origin of the graph but gives a finite field value which shows that it is not an ordinary spin-flop transition. To investigate this transition in more detail in particular in the dynamic regime, we measured low-energy spin excitations in this magnetic field range. One of the HF-ESR transmission spectra, measured at $T = 4$~K and a frequency of $f = 708$~GHz, is shown in Fig.~\ref{mag}b. The spectrum features four absorption peaks. The two most prominent resonances $\omega_1$ and $\omega_\mathrm{SF}$ at $25.6$~T and $37.4$~T are almost symmetric to the spin-flop transition and show similar intensity. An additional sharp but less intense resonance $\omega_\mathrm{DM}$ is observed slightly below the spin-flop field at $B\sim29.3$~T. The fourth peak $\omega_\mathrm{lin}$ appears at $B_\mathrm{lin} = 8.1$~T. It shows only weak intensity and belongs to a linear-in-field resonance branch which has an extrapolated zero-field gap of $\sim 926$~GHz and an effective $g$-value of $g_\mathrm{lin} = 1.94$. By removing a thin surface layer of the sample the relative intensity of this resonance as compared to the other ones weakens significantly which indicates that it is associated with surface defects.  

\begin{figure}	
\includegraphics[width=1\columnwidth,clip] {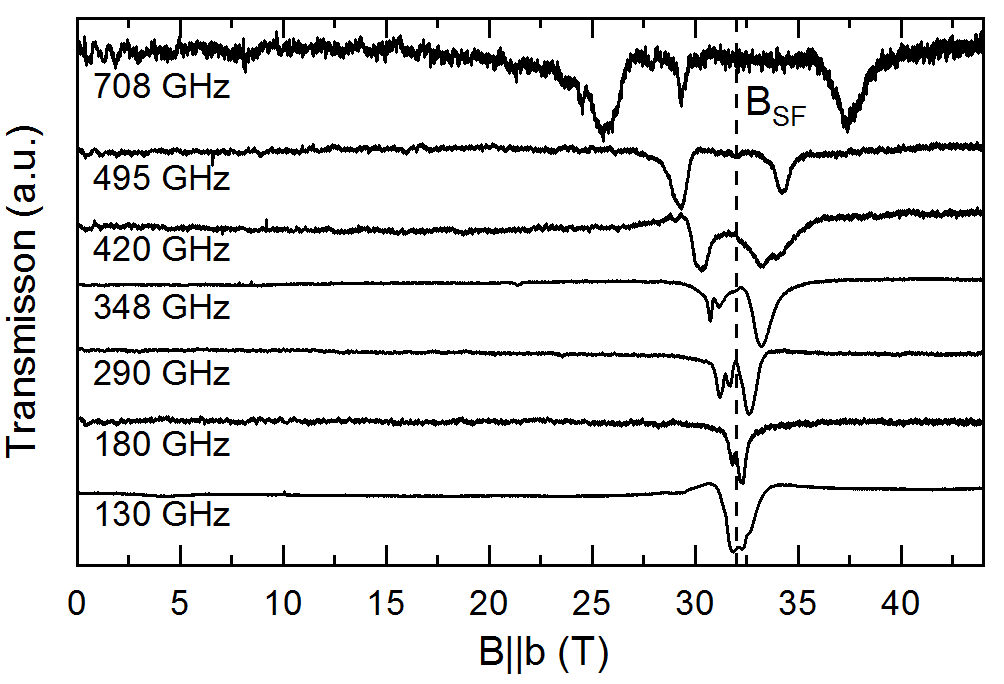}
\caption{Pulsed-field electron spin resonance up-sweep transmission spectra at $T \sim 4$~K in the frequency range between $f = 130$~GHz and $f = 708$~GHz. The vertical dashed line indicates the spin-flop field $B_\mathrm{SF}$.}\label{spectraF}
\end{figure}

From inelastic neutron scattering it is known that the magnon excitation gaps at zero field in the Brillouin zone center amount to $\Delta_1 = 1450$~GHz and $\Delta_2 = 2070$~GHz~\cite{li2006spin,toft2015anomalous,yiu2017hybrid}. The fact that we observe resonances at $708$~GHz as shown in Fig.~\ref{mag}b hence already implies the presence of a resonance branch which is softened by fields $B\|b$-axis. Further progression of softening can be seen in Fig.~\ref{spectraF} where the evolution of the resonances at lower frequencies is presented. The data show that splitting of the two main resonance features $\omega_1$ and $\omega_\mathrm{SF}$ shrinks with decreasing frequency and both peaks merge at lowest frequencies under study. At $f = 130$~GHz, only one resonance at the position of the spin-flop field is observed, which indicates complete closing of the excitation gap, i.e., complete softening of the resonance branch. This behaviour is typical for a  spin-flop transition which evolves when the external magnetic field compensates the effective field acting on the magnetic sublattice pointing in the direction opposite to the external magnetic field. In addition to softening and merging of the split resonances $\omega_1$ and $\omega_\mathrm{SF}$, the data show that the resonance $\omega_\mathrm{DM}$ completely vanishes with decreasing frequency and cannot be observed below $f \sim 700$~GHz.  

\begin{figure*}	
\includegraphics[width=1\columnwidth,clip] {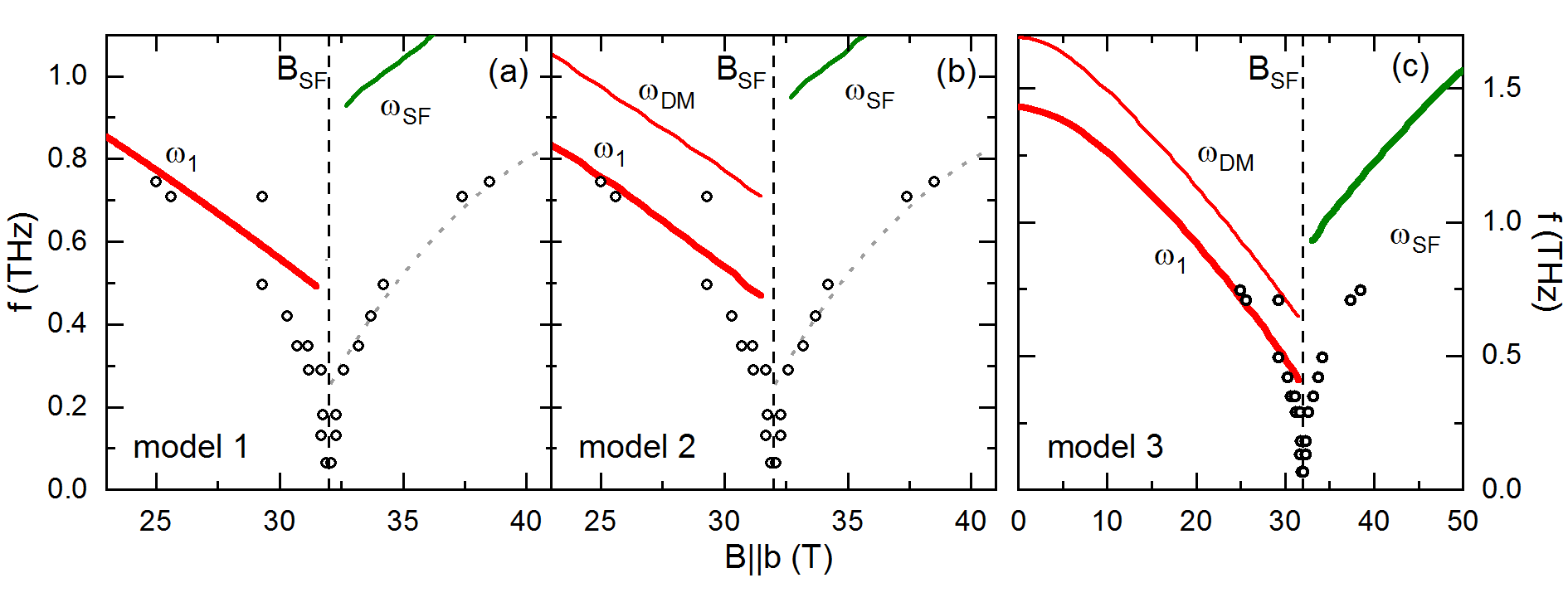}
\caption{Resonance fields at different frequencies (data points, cf. Fig.~\ref{spectraF}). Vertical dashed lines indicate the spin-flop field, gray dotted lines are guide to the eye. Solid lines are model calculations with (a) exchange interactions $J$ and anisotropy $D_\mathrm{b}$ being magnetic field-dependent parameters (Model 1), (b) with additional Dzyaloshinskii–Moriya interaction (Model 2), and (c) with a magnetic field dependent anisotropy tensor $D$ (Model 3). See Table \ref{para} for simulation parameters {\rk and Fig.~S1 (Supplemental Material) for a further comparison of the models.}}\label{FH}
\end{figure*}

Collecting the resonance fields in a field vs. frequency plot (see Fig.~\ref{FH}) evidences that the magnon gaps not only close but the resonance branches $\omega_1$ and $\omega_\mathrm{SF}$ exhibit a pronounced curvature when approaching $B_\mathrm{SF}$. 
In general, softening of the gaped antiferromagnetic resonance (AFMR) modes is expected but the very strong suppression of the resonance branch is by far not covered by AFMR models~\cite{nagamiya1951theory,nagamiya1954theory} with fixed zero-field splittings $\Delta_1$, $\Delta_2$ and spin-flop field $B_\mathrm{SF}$. 
We recall the large size of $\Delta_1$ as compared to \bsf . Specifically, the standard AFMR model would result in the effective $g$-factor of $g = 3.3(1)$ which is unreasonably large and does not match to $g_b = 2.31$ obtained from analysis of high-temperature magnetic susceptibility data~\cite{werner2019high}. Hence, one has to conclude that parameters obtained at zero magnetic field, i.e., single-ion anisotropy $D$ and exchange interaction $J$, do not describe the experimental results obtained at high magnetic field. The presence of pronounced magnetoelastic effects and associated strong magnetostrictive length changes in \lfp~\cite{werner2019high} however already suggests that microscopic parameters are supposed to be magnetic field dependent. We attribute the apparent failure of the zero-field parameter-based model to field dependence of microscopic parameters and will present an appropriate model in the following.

The spin Hamiltonian presented below extends the one used to describe recent INS data~\cite{toft2015anomalous} by magnetic fields applied along the $b$-direction and magnetic field dependent exchange interaction and anisotropy parameters. Specifically, a four magnetic sublattice model is assumed, as indicated in Fig.~\ref{magStru}. The sublattices are coupled via the exchange interactions $J_{ac}$, $J_{ab}$ and $J_{bc}$. Note, that $J_b$ does not significantly affect the results at $B<$~\bsf\ and hence is omitted in the Hamiltonian Eq.~\ref{spinH}. Interactions along the crystallographic directions are neglected in our model, because these interactions couple one sublattice to itself. In addition to the exchange interactions we add an orthorhombic anisotropy term $D$ as well as Dzyaloshinskii–Moriya (DM) interaction between the sublattices 1-2 and 3-4. {\rk As a minimal model to describe the field dependence, for} the exchange interactions and the single-ion anisotropy we assume a quadratic magnetic field dependence of the form $J(B) = J(0)\times(1-\eta B^2)$ and $D(B) = D(0)\times(1-\iota B^2)$. The Hamiltonian for one unit cell hence reads:

\begin{align}
\mathcal{H} =& \,4J_{ac}(S_1\!\cdot\! S_2+S_3\!\cdot\! S_4)+4J_{ab}(S_1\!\cdot\! S_3+S_2\!\cdot\! S_4)\notag\\
&+4J_{bc}(S_1\!\cdot\! S_4+S_2\!\cdot\! S_3)+\sum_{i,j}^{\phantom{i}} D_i{\left(S_j^i\right)}^2\notag\\
& +J_\mathrm{DM}\cdot (S_1\!\times\! S_2 + S_3\!\times\! S_4) - g\mu_\mathrm{B}B
\label{spinH}
\end{align}

Here, $S^i (i = 1,2,3)$, is the $i$-th component of the sublattice spin $S_j$ (j = 1...4), and $J_\mathrm{DM}$ is the DM interaction.~\footnote{$S$ is the thermal average of the spin, so that the magnetisation reads $M = g\mu_\mathrm{B}S$.} The value of the $g$-factor is fixed to $g=2.31$ as determined from the high-temperature magnetic susceptibility~\cite{werner2019high}. To obtain the magnetization along the $b$-direction, firstly the spin structure is determined by numerically minimizing the Hamiltonian \ref{spinH} and then the components of the spins in $b$-direction are calculated. The undetermined parameters in this Hamiltonian are the magnetic field dependencies $\eta$, $\iota$ and the strength of the DM interaction $J_\mathrm{DM}$. The best fitting results for these free parameters to describe the experimental magnetisation data $M(B||b$ (see Fig.~\ref{mag}) are $\eta = 4.6 \times 10^{-5}$~T$^{-2}$, $\iota = 8.0 \times 10^{-5}$~T$^{-2}$, and for the DM interaction an upper boundary of $J_\mathrm{DM}\leq 0.3$~meV is obtained. {\rk Note, that including further terms to the Hamiltonian Eq.~(1) or/and assuming higher order terms in the field dependencies of the microscopic parameters will yield quantitative changes while the qualitative scenario of remains robust.} Even better fittings can be obtained if a stronger magnetic field dependence, $\eta  = 8.1\times10^{-5}$~T$^{-2}$ and $\iota = 2.6\times10^{-4}$~T$^{-2}$, for $J$ and $D_b$ are assumed.~\footnote{The resulting field dependencies quantitatively depend on the actual microscopic parameters at $B=0$~T which are taken from Ref.~\cite{toft2015anomalous}.} However such a strong magnetic field dependence would lead to a change of the easy axis at high fields which is not indicated by the data. We hence restricted the parameter range to small changes of $J$ and $D$ with magnetic fields while excluding strong changes of spin orientation and spin structure. It should be noted, that small DM interactions do not change the spin configuration of the lower-field phase $B<B_{\rm SF}$. While DM interactions tempt to cant spins of sublattices 1 and 2 with respect to each other, the presence of large anisotropy suppresses canting. In contrast, dynamic response is changed drastically by DM interactions as additional resonances appear.

To simulate the excitations of Hamiltonian \ref{spinH}, the Landau–Lifshitz–Gilbert equation (the spin equation of motion)
\begin{align}
\frac{\operatorname{d}\!S_j}{\operatorname{dt}} = -\frac{\gamma}{1+\alpha^2}\left(S_j\times B_{\mathrm{eff,}j} + \alpha S_j \times ( S_j \times B_{\mathrm{eff,}j})\right) \label{eom}
\end{align}
is solved~\cite{hillebrands2003spin}. It includes the gyromagnetic factor $\gamma$ and the effective magnetic fields $B_{\mathrm{eff,}j}$ acting on the spins of the $j$-th sublattice. The effective field is composed of the static mean-field and the microwave field. Damping is considered by the parameter $\alpha$. For the calculations, we chose $\alpha$ sufficiently large to solve the differential equation, but small enough to leave the position of the resonances unchanged within the resolution of the simulation. To match our experiment, the equation of motion (Eq.~\ref{eom}) is solved for unpolarized microwaves in Faraday-geometry. To generate the resonance field vs. frequency diagram, constant frequency cuts are generated and the maxima in absorption are taken as resonance points which reflects the experimental field sweeps. 

\begin{table}
	\centering
     \begin{tabularx}{\columnwidth}{lXXX}
\vspace{0.2cm}						  & \textbf{Model 1} 			& \textbf{Model 2} 		& \textbf{Model 3} \\ 
$\bm{D_{a}}$\hphantom{XX} 		      &	$0.62\times r_2(B)$		    &$0.62\times r_2(B)$	    &$0.62\times r_2(B)$\\
$\bm{D_{c}}$				  		  & 1.56 						&1.56		            &$1.56\times r_3(B)$\\
$\bm{J_{bc}}$				  		  &$0.77\times r_1(B)$			&$0.77\times r_1(B)$		&$0.77\times r_1(B)$\\
$\bm{J_{ab}}$				  		  &$0.14\times r_1(B)$			&$0.14\times r_1(B)$		&$0.14\times r_1(B)$\\
$\bm{J_{ac}}$				  		  &$0.05\times r_1(B)$			&$0.05\times r_1(B)$		&$0.05\times r_1(B)$\\
$\bm{J_\mathrm{DM}}$				  & 0						    & 0.15					& 0.15
	\end{tabularx}
    \caption{Simulation parameters in units of meV for the three used models. The magnetic field dependence is marked by $r_{2,3}(B)$ and $r_1(B)$ for the anisotropy and exchange interaction, respectively. The magnetic field dependencies are given by $r_1 = 1-4.6 \times 10^{-5} B^2/\mathrm{T}^2$, $r_2 = 1-8.0 \times 10^{-5} B^2/\mathrm{T}^2$ and $r_3 = 1-1.7 \times 10^{-4} B^2/\mathrm{T}^2$.}
    \label{para}
\end{table}

A simulation of the system, with the best fit parameters of the magnetization measurement and neglecting DM interactions (model 1 in Table~\ref{para}) yields the resonance branches, $\omega_{\rm 1}$ and $\omega_{\rm SF}$, shown in Fig.~\ref{FH}a. While the simulation describes the resonance branch $\omega_1$ below \bsf\ well, it fails to reproduce its behavior in the direct vicinity of the critical field. It also completely misses the presence of the resonance $\omega_\mathrm{DM}$ and fails to reproduce the resonance branch $\omega_\mathrm{SF}$. Note, that gaps of the resonance branches around $B_\mathrm{SF}$ are due to sizable mean-fields, which are mainly caused by the fact that the magnetocrystalline anisotropy is large compared to the exchange interactions. In particular, the experimentally observed absence of excitation gaps at $B_{\rm SF}$ is inconsistent with the spin Hamiltonian parameters determined from INS at zero magnetic field. In order to cover the resonance $\omega_\mathrm{DM}$ by the model, DM interaction $J_\mathrm{DM}$ is introduced, which has been previously suggested to be of finite but small size in zero magnetic field and was hence neglected in the simulation of the spin dynamics probed by INS~\cite{toft2015anomalous,yiu2017hybrid}. The best simulation, which takes into account the position of $\omega_\mathrm{DM}$ as well as the ratio of intensities of $\omega_1$ and $\omega_\mathrm{DM}$ is achieved by $J_\mathrm{DM} = 0.15$~meV (see Table~\ref{para} model 2). As shown in Fig.~\ref{FH}b, considering the DM term indeed yields both the resonance branches $\omega_1$ and $\omega_\mathrm{DM}$. Similar to model 1, the simulated results however differ from the experimental data in the vicinity of \bsf . In the flopped phase no additional resonance is observed, however the resonance branch $\omega_\mathrm{SF}$ is slightly shifted to higher energies as compared to model 1. 

In order to further improve the model by employing field dependent microscopic parameters, we introduce magnetic field dependence of $D_c$ (see Table~\ref{para} model 3). Magnetic field $B||b$-dependence of $D_c$ with $D_c(B)>D_a(B)$ does not change the simulated magnetisation shown in \ref{mag}, because for small DM interactions and $B||b$ the spins are confined to the ab-plane. In addition, also $\omega_1$ and $\omega_{\rm DM}$ are only affected insignificantly for fields well below \bsf . While, there are clear changes around the spin-flop transition leading a good description of the modes in this field range, too. We obtain the magnetic field dependence $\iota_2 = 1.7 \times 10^{-4}$~T$^{-2}$ of $D_c(B)$ which was derived by restricting $D_a(B)<D_c(B)$ for all magnetic fields. 

\begin{figure}	
\includegraphics[width=1\columnwidth,clip] {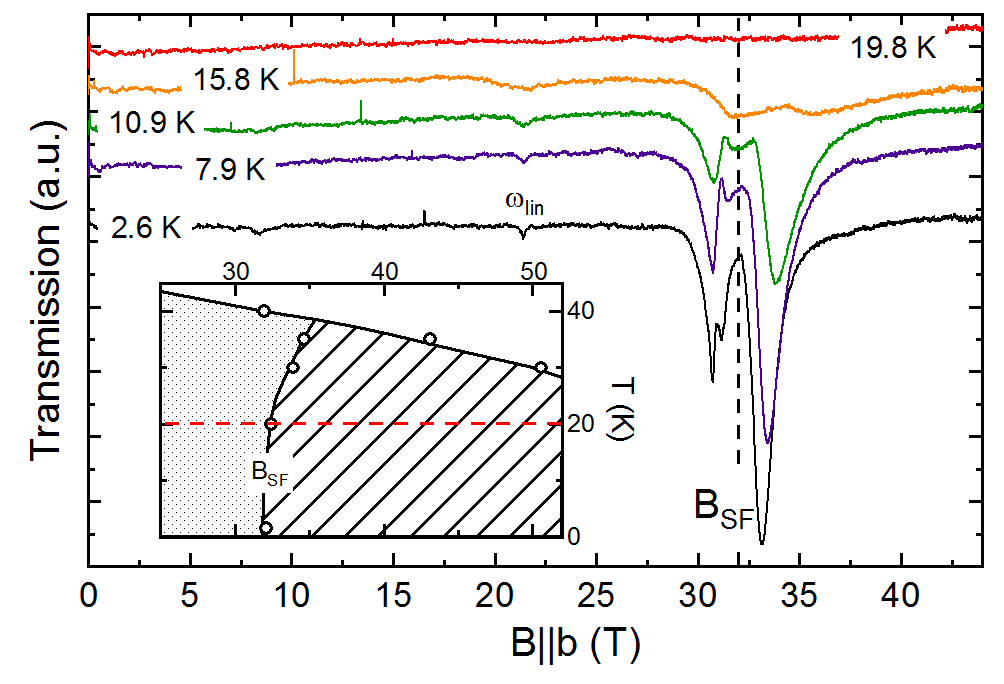}
\caption{HF-ESR transmission spectra measured at $f = 348$~GHz in the temperature regime between $2.6$~K and $\sim 20$~K. The vertical dashed line indicates the spin-flop field. Inset shows the magnetic phase diagram of \lfp, data from Ref.~[\onlinecite{werner2019high}]. The horizontal dashed (red) line marks the position, where the $T = 19.8$~K spectrum was measured.}\label{spectraT}
\end{figure}

Transmission spectra at a fixed frequency, presented in Fig.~\ref{spectraT}, show the dependence of the resonances discussed above upon increasing the temperature. At $T = 2.6$~K the resonances $\omega_1$ and $\omega_\mathrm{SF}$ are visible. The resonance $\omega_1$ shows an additional splitting at this temperature. With increasing temperature the intensity of the peaks diminishes. While the position and half width of $\omega_1$ do not significantly change upon heating, the resonance $\omega_\mathrm{SF}$ shifts with temperature to higher fields and broadens significantly. At $T\approx 16$~K the resonance $\omega_1$ has completely vanished while $\omega_{\rm SF}$ remains as a broad feature similar to a feature at \bsf . At $T = 20$~K, no resonance is observed up to $B = 45$~T. 

\section{Discussion}

The experimental data are well described by the used models for magnetic field $B\leq B_\mathrm{SF}$. In particular, the presence of sizable DM interaction is evidenced by the presence of $\omega_\mathrm{DM}$. Introducing magnetic field-dependent parameters yields an excellent description of the field dependence of AFMR modes and resolves the quantitative contradiction between zero-field gap as detected by INS and the actual value of \bsf . {\rk To be specific, the extended model shown in Fig.~\ref{FH} reproduces both the experimentally observed $\omega_1$ and $\omega_{\rm DM}$ resonance features and the bending on the former when approaching \bsf .} The field-dependence of microscopic parameters is motivated by the presence of pronounced magnetostrictive effects in \lfp ~\cite{werner2019high}. While the presence of DM interaction was suggested recently, it has neither been directly evidenced experimentally yet nor quantified in the dynamic response probed by INS~\cite{toft2015anomalous,yiu2017hybrid}. Analysis of our high-field ESR data yields $J_\mathrm{DM}\approx 150~\mu$eV. Note, that this value can not straightforwardly extrapolated to zero magnetic field as our analysis implies the above mentioned field dependence of spin Hamiltonian parameters. In contrast to excellent description of the AFMR modes below \bsf , all applied models fail to quantitatively reproduce the observed resonances above $B_\mathrm{SF}$. One may speculate that the spin structure in the high-field phase is not a purely reoriented one but further differs from the ground state. In addition, the spin-flop transition might be accompanied by structural distortions. Such distortions could change the anisotropy or the exchange interactions not in the continuous manner, assumed here, but rather abruptly. This is somehow corroborated by the fact that the spectra presented in Fig.~\ref{spectraT} indeed show that the resonances below and above \bsf\ are clearly different regarding the spectral weight. In addition, while, at $f=348$~GHz, $\omega_1$ does not shift upon heating, there is a clear temperature dependence of $\omega_\mathrm{SF}$ which shifts to higher fields. This suggests a decrease of internal fields upon heating while for $B<B_{\rm SF}$ constant internal mean-fields are evidenced. Notably, in contrast to our experimental findings, conventional mean-field theory would suggest a shift of $\omega_\mathrm{SF}$ to lower fields upon heating~\cite{nagamiya1954theory}. We also note that rather large modulations of exchange interactions or anisotropy at phase boundaries and resulting changes in the magnon spectra have been previously observed by HF-ESR in other compounds, e.g. in HgCr$_2$O$_4$\cite{kimura2011large}, CuFeO$_2$ \cite{kimura2011multifrequency}, CdCr$_2$O$_4$ \cite{kimura2015evolution}, or Cu(pz)$_2$(ClO$_4$)$_2$ \cite{Povarov}. 

The disappearance of the resonances at $T = 20$~K is somehow surprising, because the measurement takes place well in the long-range antiferromagnetically ordered phase as illustrated by the dashed red line in the phase diagram (see the inset of Fig.~\ref{spectraT}). Specifically, the phase diagram indicates no phase transition or any feature separating a region where AFMR modes are detected in the vicinity of \bsf\ (i.e., below the dashed line in the inset of Fig.~\ref{spectraT}) from the one without observable features. We recall that the antiferromagnetic order parameter in \lfp\ is only very weakly temperature dependent up to $20$~K and it only marginally depends on small fields in related materials \lmp\ and \lcp ~\cite{li2006spin,toft2015anomalous,tian2010neutron,toft2012magnetic,fogh2017magnetic}. We also note that there are only very small changes of the magnetisation between $1.5$~K and $20$~K at any given field up to $B = 50$~T, i.e., the uniform susceptibility $\chi(\omega = 0) = \partial M/\operatorname{\partial B}$ is rather constant in this temperature and field range. Note, however, that the slope of the phase boundary \bsf ($T$) starts to increase above 20~K which may indicate that the presence of fluctuations or of competing interactions is associated with the disappearance of the AFMR resonances. From an experimental point of view, there are two obvious reasons which can lead to no observable resonance at $f = 348$~GHz. These are (1) enormous broadening, leading to resonances indistinguishable from the background and (2) shift of the resonance branches above the measured frequency regime. Following the tendency of spectra up to $16$~K, both options are very unlikely. Alternatively, one may consider excitations which change the length of the spin, which is in our model constant. In Ref.~\onlinecite{yiu2017hybrid} such resonances, there called hybrid excitations, are found by inelastic neutron scattering at $\Delta = 1088$~GHz in zero magnetic field. If such excitations exist, a spectral shift of intensity from the conventional magnon excitations to spin stretching modes could take place at higher temperatures around the spin-flop transition. However, our ESR data do not show any sign of such hybrid excitation. 

In order to further clarify the magnetic field dependence of anisotropy parameters and DM interaction, measurements in the terahertz regime will be needed. We hence suggest to study the resonance branch $\omega_2$ which exhibits zero field splitting of $\Delta_2 = 2070$~GHz~\cite{yiu2017hybrid}, as this branch is particularly sensitive to changes of $D_c$. While, below $\approx 25$~T, $\omega_1$ only weakly depends on $D_c$ as shown in Fig.~\ref{FH}c, the slope of $\omega_2$ is significantly affected. Recent THz-absorption spectroscopy data on other olivine-structured phosphates show divers behavior. In \lcp , two modes are observed in zero magnetic fields which split into four modes upon application of magnetic field along the easy axis, two of which being suppressed and two stabilised in the fields~\cite{kocsis2018identification,kocsis2019magnetoelectric}. All four branches show rather linear field dependence of similar absolute slope, i.e., similar effective $g$-factor. In contrast, low-lying excitation branches in \lnp\ are non-linear already at relatively small magnetic fields applied along the easy magnetic axis~\cite{peedu2019spin}.

We finally note that our bare experimental data do not unambiguously tell whether the observed modes are classical antiferromagnetic resonances, i.e., whether they are only magnetic dipole active, or whether they are electromagnons and can be excited by oscillating electrical fields, too. As illustrated, e.g., by the example of multiferroic TbMnO$_3$, electromagnons and antiferromagnetic resonances can coexist~\cite{pimenov2009magnetic} which in principle cannot be excluded for \lfp\ in magnetic fields, either. However, the magnetic structure of \lfp\ implies at least two AFMR branches which presence at zero field was confirmed by INS. Attributing $\omega_1$ to AFMR is further corroborated by the fact that $\omega_1$ follows the magnetisation which is expected for an antiferromagnetic resonance. In additon, the magnetoelectric effect is comparably small in \lfp\ and there is no electric order which further motivates attributing the observed modes to the expected AFMR ones. 

\subsection{Summary}

We report antiferromagnetic magnon excitations in \lfp\ in high fields around the spin flop transition. The data reveal closing of the energy gap exactly at \bsf\ and hardening the AFMR modes for higher fields. An additional mode below \bsf\ is assigned to DM interaction which size is estimated to $150\mu$eV. The mean-field theory model obtained from previous zero-field INS data only describes the field dependence of the magnon modes and of the magnetisation if exchange interactions and magnetocrystalline anisotropy are considered to change with magnetic field. This also holds for the value of the spin-flop field. The AFMR modes disappear at around 20~K, i.e., well in the long-range AFM ordered phase, which might suggest the presence of spin fluctuations or competing interactions.

\begin{acknowledgements}
The project is supported by Deutsche Forschungsgemeinschaft (DFG) through KL 1824/13-1 and the Heidelberg STRUCTURES Excellence Cluster (EXC2181/1-390900948). Funding by BMBF via project SpinFun (project 13XP5088) is gratefully acknowledged. We acknowledge the support of the HLD at HZDR, member of the European Magnetic Field Laboratory (EMFL). S.Z. and A.P. acknowledge the support of DFG  through ZV 6/2-2.
\end{acknowledgements}

\bibliography{Bib_LiFePO4}

\end{document}